\documentclass[aps,pra, groupedaddress,superscriptaddress]{revtex4-2}
\usepackage{graphicx} % for including figures
\usepackage{amsmath,amssymb}
\usepackage{listings}
\usepackage{xcolor} % for custom colors

\newcounter{code}
\renewcommand{\thecode}{\arabic{code}}

\newcommand{\codecaption}[1]{
    \refstepcounter{code}
    \par\vspace{\baselineskip}
    \par\noindent\textbf{Code \thecode:} #1
    \par\vspace{\baselineskip}
}

\usepackage[utf8x]{inputenc}
\usepackage{array}
\usepackage{textgreek}
\usepackage[shortlabels]{enumitem}
\begin{document}

\title{PyMembrane: A flexible framework for efficient simulations of elastic and liquid membranes}

\author{D. A. Matoz-Fernandez}
\email{dmatoz@ucm.es}
\affiliation{Department of Theoretical Physics, Complutense University of Madrid, Madrid, 28040, Spain}

\author{Siyu Li}
\affiliation{Department of Materials Science and Engineering, Northwestern University, Evanston, 60208, United States}
\affiliation{Department of Physics and Astronomy, Northwestern University, Evanston, 60208, United States}

\author{Monica Olvera de la Cruz}
\affiliation{Department of Materials Science and Engineering, Northwestern University, Evanston, 60208, United States}
\affiliation{Department of Physics and Astronomy, Northwestern University, Evanston, 60208, United States}
\affiliation{Department of Chemistry, Northwestern University, Evanston, 60208, United States}

\author{Rastko Sknepnek}
\email{r.sknepnek@dundee.ac.uk}
\affiliation{School of Science and Engineering, University of Dundee, Dundee, DD1 4HN, United Kingdom}
\affiliation{School of Life Sciences, University of Dundee, Dundee, DD1 5EH, United Kingdom}

\begin{abstract}
PyMembrane is a software package for simulating liquid and elastic membranes using a discretisation of the continuum description based on unstructured triangulated two-dimensional meshes embedded in three-dimensional space. The package is written in C++, with a flexible and intuitive Python interface, allowing for a quick setup, execution and analysis of complex simulations. PyMembrane follows modern software engineering principles and features a modular design that allows for straightforward implementation of custom extensions while ensuring consistency and enabling inexpensive maintenance. A hallmark feature of this design is the use of a standardized C++ interface which streamlines adding new functionalities. Furthermore, PyMembrane uses data structures optimised for unstructured meshes, ensuring efficient mesh operations and force calculations. By providing several templates for typical simulations supplemented by extensive documentation, the users can seamlessly set up and run research-level simulations and extend the package to integrate additional features, underscoring PyMembrane's commitment to user-centric design.
\end{abstract}

\maketitle
%% \linenumbers

\section{Introduction}

Membranes are ubiquitous in the living world. For example, they serve to separate cells from the environment and compartmentalise the cell's interior into various organelles. Biomembranes can selectively pass different ions and molecules thus allowing the maintenance of concentration gradients essential for the proper function of the cell, transmit signals, host biosynthetic activities, etc.~\cite{gennis2013biomembranes}. In addition to their important biological roles, membranes exhibit rich physical properties~\cite{evans2018mechanics}. This is a consequence of their quasi-twodimensional structure. 

A typical biomembrane is approximately 5 nm thick but can span several tens of \textmu m in lateral directions, and can either be fluid or elastic in nature. Fluid membranes undergo flow when subjected to shear deformation, while elastic ones do not. Due to their high surface-to-thickness ratio, membranes are flexible and can deform to form highly curved structures. It has been increasingly recognized that curvature plays an important role in biology~\cite{schamberger2023curvature}, refuelling the interest in understanding membrane physics. In contrast to polymers, however, where bending and stretching deformations are independent due to their one-dimensional structure, this is not true for membranes, resulting in unique effects on their mechanical properties~\cite{nelson2004statistical}. These all lead to very rich non-linear behaviour that requires the use of numerical simulations to investigate the physical properties of membranes. 

Biological membranes have a complex structure. They typically comprise a lipid bilayer consisting of lipids of different types interspersed with membrane and trans-membrane proteins, sugars, etc.~\cite{Alberts2002}. Therefore, modelling a specific biological function usually requires an atomistic level of detail~\cite{enkavi2019multiscale}. Such simulations are commonly used, for example, in studying processes such as transport through ion channels~\cite{kutzner2011computational}, but are typically limited to several tens of nm in size and several tens of \textmu s in length. This is, however, insufficient to probe processes that occur at much larger length and time scales, such as shape changes of the entire cell. Many coarse-graining schemes have, therefore, been introduced that allow accessing those scales at the expense of reducing the level of details, such as those proposed in Refs.~\cite{drouffe1991computer,cooke2005tunable,marrink2007martini} and reviewed in depth in Ref.~\cite{venturoli2006mesoscopic}.  
 
Although recent advances in multiscale modelling allow simulations of realistic membrane shapes~\cite{pezeshkian2020backmapping,pezeshkian2021simulating}, such models require extensive computer resources and are often unnecessarily detailed for the problem of interest. In such situations, it is advantageous to treat the membrane as a macroscopic object and model it using continuum approaches. The membrane shape is described as a smooth two-dimensional surface embedded in three-dimensional space. One assigns energy to the shape and either seeks configurations that minimise the energy or studies the dynamics by solving the equations of motion where forces are determined as negative gradients of the energy with respect to the shape. A powerful numerical strategy is to approximate the continuous two-dimensional surface using a discrete triangulated surface. An additional advantage is that once the relation between shape and energy has been established, the same method can be used to study the mechanics of other quasi-twodimensional structures that are not necessarily related to lipid membranes, or even biological.

%The aim of this work is to present a flexible and efficient implementation of a class of models based on triangulated surfaces for studying static and dynamic mechanical properties of membranes.

\subsection{Software ecosystem}

Communities working on problems that require atomistic or coarse-grained particle-based simulations have had for several decades at their disposal a number of powerful, well-documented and supported open-source and commercial packages. For example, GROMACS~\cite{gromacs}, AMBER~\cite{amber}, CHARMM~\cite{charmm}, and NAMD~\cite{namd} have been widely used for sophisticated biomolecular simulations at the atomistic scale. Packages such as ESPResSo~\cite{espresso}, LAMMPS~\cite{LAMMPS}, and HOOMD-Blue~\cite{anderson2020hoomd} are excellent tools for coarse-grained simulations. Similarly, there are numerous sophisticated tools for finite-element simulations of continuum models of fluid and solid mechanics (e.g., FEniCS~\cite{fenics}, OpenFOAM~\cite{openfoam}, FreeFEM++~\cite{fempp}, COMSOL, Abaqus~\cite{abaqus}, etc.). The triangulated membrane models, however, fall between those two categories. While it is possible to use packages such as LAMMPS and HOOMD-Blue to simulate triangulated elastic meshes~\cite{bowick2017non,russell2017stiffening,hanakata2022anomalous, chen2022spontaneous}, those tools are not primarily designed for such purposes and often do not provide the desired level of flexibility. Furthermore, these packages cannot be directly used to study liquid membranes and an entirely different set of tools is required. Most works (e.g., see Refs.~ \cite{anagnostopoulos1993fluid,Gompper1996,gompper1997network,vsiber2006buckling,katifori2010foldable,ramakrishnan2010monte,vernizzi2011platonic,sknepnek2011buckling,sknepnek2012buckling,vsaric2012fluid,davidovitch2019geometrically,Matoz2020}), therefore, typically rely on custom codes that are rarely made public.

A notable exception is the Surface Evolver~\cite{brakke1992surface}. With an impressive library of surface energy models, Surface Evolver has been the main tool for simulating triangulated surfaces. Simulations with Surface Evelover are, however, limited to energy minimisation, subject to various constraints, and it is not straightforward to study dynamics. Furthermore, the original code base, written in ANSI-C, is dated, and no longer under active development, making it hard to modify and extend. 

Recently, three new packages have emerged, TriMem~\cite{Siggel2022trimem}, flippy~\cite{dadunashvili2023flippy}, and FreeDTS \cite{pezeshkian2023FreeDTS}, designed for Monte Carlo simulations of triangulated models of lipid membranes. These packages use modern software design, are well-documented and straightforward to use. TriMem has been parallelised making it capable of simulating large systems. 

The purpose of this work is to add to this growing ecosystem a powerful, yet easy-to-use package framework for stimulating a wide class of triangulated surface models of elastic and liquid membranes that combines the efficiency of low-level implementation in C++ with the intuitive user-facing Python interface. 

\section{Physics of membranes}

In this section, we give a brief overview of the continuum models of membranes. For a detailed description, for example, see Ref.\ ~\cite{nelson2004statistical}.

\subsection{Continuum model of a membrane}

A membrane is modelled as a two-dimensional curved surface embedded in a three-dimensional flat ambient space. Points on the surface are described by a radius vector $\mathbf{r}=\mathbf{r}(s^1,s^2)=x(s^1,s^2)\hat{\mathbf{i}} + y(s^1,s^2)\hat{\mathbf{j}} + z(s^1,s^2)\hat{\mathbf{k}}$, where $s^1$ and $s^2$ are the intrinsic (i.e., curvilinear coordinates), and $\hat{\mathbf{i}}$, $\hat{\mathbf{j}}$, and $\hat{\mathbf{k}}$ are the three unit-length basis vectors of the ambient space. For example, for a sphere of radius $R$, a commonly used parametrisation is $s^1\equiv\vartheta$ and $s^2\equiv\varphi$, with $\mathbf{r}(\vartheta,\varphi)=R\sin\vartheta\cos\varphi\hat{\mathbf{i}}+R\sin\vartheta\sin\varphi\hat{\mathbf{j}}+R\cos\vartheta\hat{\mathbf{k}}$. One defines two tangent vectors to the surface, $\mathbf{e}_\alpha = \partial_\alpha\mathbf{r}$, where $\alpha\in\{1,2\}$ and $\partial_\alpha \equiv \frac{\partial}{\partial s^\alpha}$. In general, tangent vectors are neither orthogonal to each other nor of unit length. Tangent vectors define the unit normal vector to the surface, $\mathbf{n}=\left(\mathbf{e}_1\times\mathbf{e}_2\right)/\left|\mathbf{e}_1\times\mathbf{e}_2\right|$, and the rank $(0,2)$ metric tensor $\hat{g}$ with components   $g_{\alpha\beta}=\mathbf{e}_\alpha\cdot\mathbf{e}_\beta$. Armed with these quantities, one defines the local rank $(1,1)$ curvature tensor with components,
\begin{equation}
    C_\alpha^\beta = -g^{\beta\gamma}\mathbf{e}_\alpha\cdot\partial_\gamma\mathbf{n},
\end{equation}
where $g^{\beta\gamma}$ are components of a rank $(2,0)$ tensor that is the inverse of the metric tensor, and summation over pairs of repeated indices is assumed. $C$ is, therefore, just a $2\times2$ symmetric matrix with two real eigenvalues, $\lambda_1$ and $\lambda_2$, called principal curvatures. One defines the mean curvature, $H=\frac{1}{2}\left(\lambda_1+\lambda_2\right)$, and the Gaussian curvature, $K=\lambda_1\lambda_2$. For a sphere of radius $R$, $H=1/R$ and $K=1/R^2$ everywhere. For an arbitrary surface, $H$ and $K$ are position-dependent and can be both positive (i.e., bowl-like) and negative (i.e., saddle-like). 

The energy penalty for bending deformations is then given in terms $H$ and $K$ as~\cite{helfrich1973elastic},
\begin{equation}
    \mathcal{E}_\text{bend} = \int_M\text{d}A\left[\frac{\kappa}{2}\left(H-H_0\right)^2 + \bar{\kappa}K\right],\label{eq:helfrich}
\end{equation}
where $H_0$ is the spontaneous curvature, $\kappa$ is the bending modulus, $\bar{\kappa}$ is the saddle-splay modulus, and $\int_M\text{d}A$ indicates that the integral is over the two-dimensional surface of the membrane, with $\text{d}A=\sqrt{\det{g}}\text{d}s^1\text{d}s^2$. Eq.~\eqref{eq:helfrich} describes low-energy deformations of a liquid membrane. 

For an elastic membrane, one needs to include the energy penalty of stretching deformations, which are, in the linear response regime, given as~\cite{pomeau2010elasticity},
\begin{equation}
    \mathcal{E}_\text{stretch} = \int_M\text{d}A \mathcal{A}^{\alpha\beta\gamma\delta}u_{\alpha\beta}u_{\gamma\delta},\label{eq:stretch}
\end{equation}
where $\mathcal{A}^{\alpha\beta\gamma\delta}$ components of the rank $(4,0)$ elastic tensor that encodes elastic properties of the membrane and $u_{\alpha\beta}$ are components of the rank $(0,2)$ stain tensor that quantifies the amount of stretching. For an isotropic material, the elastic tensor has only two independent components, the two Lam{\' e} coefficients. 

The total energy can also be augmented to include various constraints (e.g., total volume). These can either be hard (i.e., imposed by Lagrange multipliers) or soft (i.e., imposed as soft harmonic potentials~\cite{vsiber2006buckling}). Finally, if the membrane has a boundary, one can add various boundary terms to the energy (e.g., line tension). 

\subsection{Triangulated surfaces}

In order to perform numerical simulations, the surface needs to be discretized. This means the membrane is represented as a triangulated mesh, i.e., a piece-wise linear approximation of its surface. The mesh consists of vertices connected by edges such that three edges form a triangle. Edges cannot cross, triangles cannot overlap, and there can't be any ``dangling" vertices or edges (i.e., those not belonging to a triangle). The entire surface must be covered by the mesh. The mesh can either have a boundary, form a closed surface, or be subject to periodic boundary conditions.

It is also necessary to construct discrete analogues, $E_\text{bend}$ and $E_\text{stretch}$, of the continuum bending and stretching energies defined in Eq.~\eqref{eq:helfrich} and Eq.~\eqref{eq:stretch}~\cite{bian2020bending}, as well as discrete versions of boundary energy terms, and constraints. There are many ways to do this, with various levels of tradeoffs between accuracy and efficiency~\cite{brakke1992surface}. PyMembrane implements multiple such models. Some of the examples will be discussed in Section~\ref{sec:examples}, with the full list of available models given in the accompanying documentation. 

For simulations of fluid membranes, vertices have to be allowed to diffuse. This is achieved, e.g., by allowing a bond flip, i.e., an edge shared by two triangles is removed and the two vertices opposite to it are reattached by a new edge. Mesh connectivity, hence, becomes a dynamic variable~\cite{gompper2004triangulated}. 

Membrane conformation is updated either stochastically, by a Monte Carlo procedure, or dynamically, by integrating equations of motion. The Monte Carlo approach involves moving vertices and, in the case of fluid membranes, flipping edges at random and accepting or rejecting the move, e.g., based on the Metropolis rules. The dynamic approach involves finding forces on each vertex by computing the gradient of the total energy with respect to the vertex position, $-\nabla_{\mathbf{r}_i}E$, and using it to integrate either first (i.e., overdamped) or second order (i.e., Newton) equations of motion. In the case of integrating Newton's equations of motion, mass is typically assigned to vertices. Sometimes, it is convenient to combine both Monte Carlo and dynamics approaches. PyMembrane, therefore, implements them both allowing for flexible hybrid simulations.

\section{Design and implementation}

% Place figure captions after the first paragraph in which they are cited.
\begin{figure*}[t]
  \centering
  \includegraphics[width=0.9\textwidth]{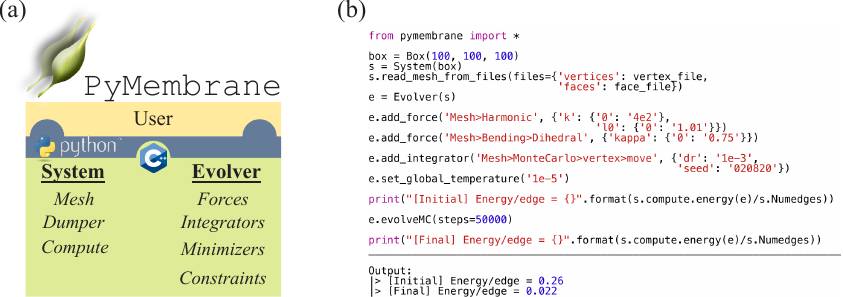}
  \caption{{\bf PyMembrane design and implementation.} (a) The C++ back-end consists of two main components, \emph{System} and \emph{Evolver}, implemented as C++ classes and exposed to users via a Python interface provided by pybind11~\cite{pybind11}. The \emph{System} class handles the simulated system by tying together five components, implemented as separate classes, three classes that handle the mesh (vertices, edges, and triangles), two auxiliary classes to compute geometrical properties (e.g., area, volume, etc.) of the mesh, and one class that handles the output in formats suitable for visualization and analysis. The \emph{Evolver} class handles the dynamics of the system by keeping track of the list of Forces, Integrators, Minimizers, and Constraints. (b) A basic example showing the simplicity and flexibility of the PyMembrane Python interface in action.}
  \label{fig1}
  \end{figure*}

\subsection{Design principles}

PyMembrane is a software package for numerical simulations of liquid and elastic membranes using the triangulated mesh representation. For efficiency, the backend is implemented in C++ and exposed to users as a set of Python classes with pybind11~\cite{pybind11}. This creates an intuitive, easy-to-use, yet powerful interface and thus combines the expressiveness of Python with the speed of C++. It allows setup, execution, analysis, and visualisation of complex, research-level simulations without the need for understanding low-level implementation details. The development, maintenance, and possible expansion of the code base are guided by modern software engineering principles~\cite{martin2017}.  

Recognizing the diverse research requirements, the design strategy emphasizes modularity and extensibility. As illustrated in fig.~\ref{fig1}a, PyMembrane consists of a number of loosely coupled modules that operate as nearly independent entities. Each component interfaces only with the mesh and a handful of support classes, therefore allowing for straightforward ways to modify, extend, and implement new models and simulation algorithms. The code base is organised in a clear and intuitive way with detailed comments that allow for automatic documentation generation with Doxygen, simplifying further navigation of the source code. 

\subsection{Implementation overview}

\begin{table}
\centering

\caption{Overview of the available bending (b), stretching (s) and boundary (bd) energy models, Monte Carlo (MC), Brownian dynamics (BD), and Velocity Verlet (VV) integrators, and energy minimisers currently implemented in PyMembrane. \label{table:1}}
\begin{tabular}{ |p{3.5cm}|p{5cm}|p{3cm}|  }
 \hline
 \multicolumn{3}{|c|}{PyMembrane implemented methods} \\
 \hline
 models & integrators & minimizers\\
 \hline
 harmonic (s)~\cite{Seung88}   & vertex move (MC)~\cite{nelson2004statistical}    & FIRE~\cite{bitzek2006fire}\\
 Cauchy-Green (s) ~\cite{sknepnek2012nonlinear} &   vertex swap (MC)  &\\
 limit (s)~\cite{abraham1989, nelson2004statistical} & edge swap (MC) & \\
 line tension (bd)~\cite{demers2012curvature}    & edge flip (MC) \cite{kazakov1985critical}& \\
 harmonic area (s)  &   vertex dynamics (BD, VV)~\cite{snook2006langevin}  & \\
 dihedral (b)~\cite{Kantor1986,Seung88} &  &    \\
 Helfrich (b)~\cite{gompper1991fluctuations} &   & \\
 \hline
\end{tabular}
\end{table}

\begin{figure}[!h]
\centering
\includegraphics[width=0.9\textwidth]{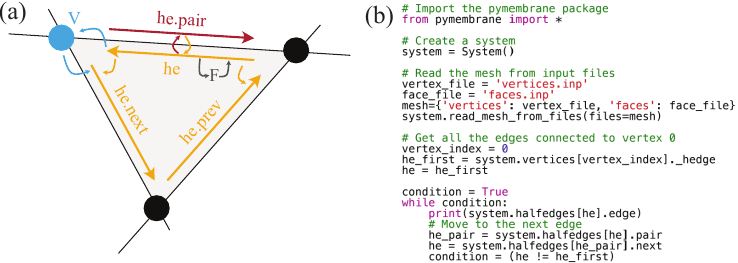}
\caption{(a) Schematic representation of the half-edge data structure with stored connectivity data indicated by arrows. (b) The Python code snippet demonstrates the connectivity information stored in the half-edge data is used to retrieve all edges connected to a particular vertex.}
\label{fig2}
\end{figure}

At the core of the PyMembrane package is the triangulated mesh representation of the membrane surface. The mesh is implemented using the half-edge data structure~\cite{weiler1985edge} that consists of four classes: vertices, half-edges, edges, and faces (i.e., triangles). Each mesh element has a list of properties (e.g., type, age, reference metric, etc.) assigned to it. Properties are passed to the mesh element classes as parameters. As sketched in fig.~\ref{fig2}a, an edge between vertices $i$ and $j$ is split into two directed half-edges, one pointing from $i$ towards $j$ and the other from $j$ towards $i$. The mesh connectivity is encoded as follows. Each half-edge contains information about the vertex it originates from and points to, the half-edge that precedes it, the one that follows it, its opposite-pointing pair, and the face to the left when looking along it. A vertex stores the information about one of the half-edges that originate at it. A face keeps track of one of the half-edges that belong to it. Within each face, half-edges are ordered counterclockwise ensuring consistent orientation of all triangles in the mesh. This layout allows for a straightforward traversal of all mesh elements, as shown in the code snippet in fig.~\ref{fig2}b. The half-edge data structure is typically implemented using pointers. While very elegant, the pointer-based implementation is not suitable for parallelisation. In PyMembrane, we have, instead, implemented it using the C++ standard library vector class, which makes it parallelisable both on CPUs and GPUs~\cite{Matoz2020}.  
% Place figure captions after the first paragraph in which they are cited.

In line with our vision of creating a versatile tool, PyMembrane provides functionalities that enable simulations of a wide range of problems. It comes with a variety of models of stretching, bending, and boundary energies, a capability to simulate multicomponent liquid and solid systems, the ability to find energy-minimising shapes as well as to study the effects of thermal fluctuations, etc. An overview of available methods is given in Table~\ref{table:1}. Each of these components is implemented as a separate class exposed to Python. This allows them to be combined in different ways thus providing users the flexibility to tailor simulations to their specific needs. 

\subsection{On units.}

PyMembrane does not use a specific set of units. All quantities are defined in their ``bare'' form, meaning values are not scaled. It is left to the user to assign physical significance to the numerical values. For instance, one could measure lengths in terms of the average edge length of the mesh, or energies in terms of the bending rigidity, but this isn't mandated. For further details, refer to examples in Refs.~ \cite{Matoz2020, li2021effect}.

\section{Showcase examples}
\label{sec:examples}
In this section, we demonstrate various applications of PyMembrane to study common problems in the physics of liquid and solid membranes. Fully commented, working Python codes, along with input files, are included in the package~\cite{matoz2023pymembrane}.

The aim of these demonstrations is not only to show PyMembrane's versatility but also to outline a typical workflow for setting up a simulation. For pedagogical reasons, all examples will follow the same layout, as indicated by comments in the code listings.

The key steps are:
\begin{enumerate}[label=\textbf{Step \arabic*},leftmargin=3.5\parindent]
    \item  Initialise the simulation box, i.e., create an instance of the \lstinline|Box| class.
    \item  Create an instance of the \lstinline|System()| class, which handles the mesh, boundary conditions, output for visualization, etc.
    \item  Create an instance of the \lstinline|Evolver()| class which handles energy and force models, constraints, integrators, and minimizers. This class handles simulation runs, be it using Monte Carlo simulations via \lstinline|evolveMC(steps)|, dynamic simulations via \lstinline|evolveMD(steps)|, or energy minimization via \lstinline|minimize()|.
\end{enumerate}

Translated to Python code, these steps are:

\codecaption{PyMembrane workflow: Typical use case for System and Evolver classes.}\label{code:code1}
\begin{lstlisting}[language=Python]
# Box
box = Box(Lx, Ly, Lz, periodic)
# System
s = System(box)
s.read_mesh_from_files(files={'vertices': vertex_file, 
                              'faces': face_file})
# Evolver
e = Evolver(s)
\end{lstlisting}

After setting up the \lstinline|Evolver()| class, one adds forces and integrators as illustrated below:

\codecaption{PyMembrane workflow: Force potentials and Integrators}\label{code:code2}
\begin{lstlisting}[language=Python]
# Add a force and integrator to the system
e.add_force('<PotentialName>', {'<Parameter1>': {'<Type_1>': '<Value_1>', ...},
                                '<Parameter2>': {'<Type_1>': '<Value_1>', ...},
                                ...})
# Add an integrator to the system
e.add_integrator('<IntegratorName>', {'<Parameter1>': '<Value1>',
                                      '<Parameter2>': '<Value2>',
                                       ...})
\end{lstlisting}

In all examples, unless stated otherwise, we use a specific set of energy and force models for stretching and bending of the membrane. The discrete stretching energy is modelled by assigning Hookean springs to the edges of the mesh~\cite{Seung88}, i.e.,
\begin{equation}
\label{Eq.HarmonicSpring}
E_\text{stretch} = \sum_e \frac{1}{2} k \left(l_e - l_0\right)^2,
\end{equation}
where $k$ is the spring constant (related to Young's modulus of the continuous membrane~\cite{Seung88}), $l_e$ is the length of edge $e$, $l_0$ is its rest length, and the sum is over all edges. In PyMembrane, this model is used by setting \lstinline|'Mesh>Harmonic'|, with parameters \lstinline|'k'| for the spring constant and \lstinline|'l0'| for the rest length.

For the discrete bending energy, we use one of the simplest models based on the dihedral angle between two triangles sharing an edge~\cite{Kantor1986,Seung88}. It is defined as,
\begin{equation}
\label{Eq.SeungNelsonBending}
E_\text{bend} =  \frac{1}{2}\kappa \sum_{e}  \left(1-\hat{\mathbf{n}}_{e}^{(1)}\cdot\hat{\mathbf{n}}_{e}^{(2)}\right),
\end{equation}
where $\kappa$ is the bending stiffness, while $\hat{\mathbf{n}}_{e}^{(1)}$ and $\hat{\mathbf{n}}_{e}^{(2)}$ are the unit normals to the two faces that share edge $e$. 

In order to prevent highly deformed triangles that can lead to numerical instabilities, PyMembrane implements a limit on the length of each edge~\cite{nelson2004statistical}. For Monte Carlo simulations, this is implemented using a potential of the form,
\begin{equation}
E =
   \begin{cases}
   0 & l_{min} < l_e < l_{max}\\
   \infty & \text{otherwise}
   \end{cases},
\end{equation}
where $l_{min}$($l_{max}$) is the minimum (maximum) edge length. The user can set the edge length limits by using \lstinline|'Mesh>Limit'|, with parameters \lstinline|'lmin'| for minimum and \lstinline|'lmax'| for maximum the edge length respectively.

We note that while other combinations of discrete stretching and bending energy models are available (as detailed in Table~\ref{table:1}), we have chosen to showcase the simplest ones in the interest of simplicity and speed.

\subsection{Buckling of a +1 disclination}

\begin{figure}[!t]
  \includegraphics[width=\columnwidth]{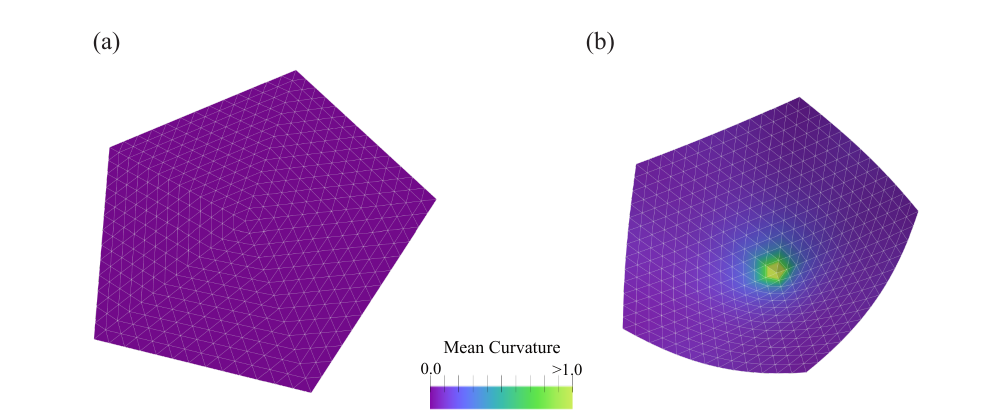}
  \caption{Snapshots of a Monte Carlo simulation of an open +1 disclination that shows buckling into a conical shape: (a) the initial flat configuration; (b) the relaxed buckled configuration. The colour bar represents the local mean curvature of the mesh.}
  \label{fig:example_1}
\end{figure}

We start with the well-known problem in thin-sheet mechanics that studies out-of-plane buckling of a +1 disclination~\cite{Seung88}. For simplicity, we discuss the underlying physics using the discrete version of the problem. Let's assume we have a hexagonal flat patch tiled by equilateral triangles. The hexagon is, therefore, made of six equilateral triangular wedges. If we remove one wedge (i.e., a section of angle $\pi/3\equiv2\pi/6$) and glue together the two free edges, the resulting plate will be a pentagon, with the vertex in the centre having five nearest neighbours. This is a +1 disclination. It is clear that removing a large part of the material and reconnecting what remains leads to substantial deformations. The elastic energy can be reduced if the patch is allowed to buckle out of the plane. This introduces some bending penalty but reduces the stretching energy. One can show that the buckling transition happens if the F\"oppl-von K\'arm\'an number $YR^2/\kappa\ge154$~\cite{Nelson87, Seung88}, where $Y$ is Young's modulus, $R$ is the patch radius, and $\kappa$ is the bending rigidity. Interestingly, this is how one makes cone-shaped party hats. Young's modulus of cardboard is orders of magnitude larger than its bending rigidity, putting the $YR^2/\kappa$ ratio in the regime where buckling is always preferred. 
By following the simple workflow detailed at the beginning of this section we can use PyMembrane to show how a +1 planar disclination can buckle out of the plane.

\codecaption{Monte Carlo simulation of the disclination problem.}\label{code:disclination}
\begin{lstlisting}[language=Python]
s = System(Box(50.0, 50.0, 50.0))
s.read_mesh_from_files(files={'vertices':'vertices_N14.dat', 
                              'faces':'faces_N14.dat'})
e = Evolver(s)
#add the potentials in this case 
# stretching, bending and limit 
e.add_force('Mesh>Harmonic', {'k':{'0': str(100.0)}, 
                                    'l0':{'0': str(1.0)}})

e.add_force('Mesh>Limit', {'lmin':{'0':str(0.7)}, 
                           'lmax':{'0':str(1.3)}})

e.add_force('Mesh>Bending>Dihedral', {'kappa':{'0':str(1.0)}})

#add the Monte Carlo integrator
e.add_integrator('Mesh>MonteCarlo>vertex>move', {'dr':'0.008'})
#set the temperature
e.set_global_temperature(str(1e-6))

## Compute the initial energy
print('[Initial] energy = {} x 10^-2'.format(s.compute.energy(e)))
#Evolve the system
e.evolveMC(steps=100000)
# Compute the final energy and dump the mesh
print('[Final] energy = {} x 10^-2'.format(s.compute.energy(e)['edges'])
s.dumper.vtk('flat_disclination')
\end{lstlisting}

Results are shown in Fig. \ref{fig:example_1}. This example serves to show how PyMembrane can be used to study open boundary conditions.

\subsection{Buckling of an icosahedral virus}

\begin{figure}[!h]
  \includegraphics[width=\columnwidth]{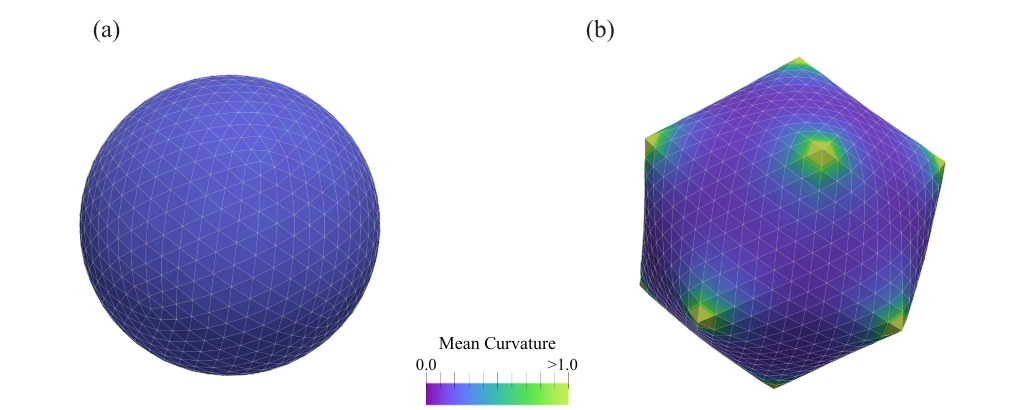}
  \caption{Snapshots of a Monte Carlo simulation of an icosahedral closed shell, representing a virus with $T=108$ ($(p,q)=(6,6)$), according to the Caspar-Klug classification~\cite{caspar1962physical}. The material parameters are chosen such that the F\"oppl-von K\'arm\'an number is $\gamma \approx 400$. (a) The sphere with 12 5-fold defects at the corners of an icosahedron prior to buckling; (b) a buckled, faceted structure. The colour bar represents the local mean curvature of the mesh.}
  \label{fig:example_2}
\end{figure}

This example is inspired by using the mechanics of thin shells to understand the shape of certain types of viruses~\cite{Lidmar03}. Caspar and Klug~\cite{caspar1962physical} showed that shell proteins in viruses form an icosahedral triangulation of a sphere formed by a set of pentavalent and hexavalent proteins.~\cite{caspar1962physical}. Due to the spherical topology, there have to be at least 12 pentavalent sites that are +1 disclinations discussed in the previous example. These disclination sites can buckle if the F\"oppl-von K\'arm\'an number is sufficiently large converting a spherical shape into a faceted polyhedron. This mechanism was used to explain why smaller viruses are spherical and larger ones are faceted~\cite{Lidmar03}. 

To demonstrate how +1 disclinations work in a closed system, one can follow the workflow outlined in Code \ref{code:disclination}. The only modifications required by the user are updating the input files and setting the spring constant to $k=350.0$. The outcome of these adjustments is illustrated in Fig. \ref{fig:example_2}. This example underscores the capability of PyMembrane in handling closed membranes.

\subsection{Energy Minimization}
\label{subsec:example1}

This example shows how to use PyMembrane to perform constant volume energy minimization of a closed elastic vesicle, i.e., a thin elastic shell. Therefore, one minimises the total elastic energy $E_\text{bend}+E_\text{stretch}$ subject to the volume constraint, i.e., one minimises $E_\lambda = E_\text{bend}+E_\text{stretch} + \lambda\left(V-V_0\right)$, where $V_0$ is the target volume and $\lambda$ is the Lagrange multiplier. The constraint is imposed as discussed in Refs.~\cite{leimkuhler2004simulating, allen2017computer}, and the volume is computed as discussed in Ref.~\cite{vsiber2006buckling}.

The first step is to create a triangulated spherical shell by using for example Gmsh~\cite{Gmsh}. Next, once that PyMembrane package is loaded into the Python environment, we create and instance of the \lstinline|System()| (see Code \ref{code:code1} and \ref{code:code2}) class and read the vertices and the faces from files. Once the mesh is loaded, we can compute quantities related to the geometry of the mesh (e.g., edge lengths and vesicle volume),

\codecaption{Computations of the mean edge length and the total volume of the vesicle.}\label{code:code3}
\begin{lstlisting}[language=Python]
mean_edge_length = mean(s.compute.edge_lengths())
print('mean edge length: {}'.format(mean_edge_length))
volume = s.compute.volume()
print('volume: {}'.format(volume))
\end{lstlisting}

As before we need to incorporate the stretching and bending potential (see, for example, Ref.~\ref{code:disclination}). We then proceed to incorporate the minimization method \cite{bitzek2006fire} using an instance of the Evolver class. 

\codecaption{Minimize.}\label{code:code4}
\begin{lstlisting}[language=Python]
e.set("Mesh>Harmonic", "l0":{"0":1.1*mean_edge_length})

fire_param =  {'dt': str(1e-2), 
              'max_iter': str(10000), 
              'etol': str(1e-3)}
e.add_minimizer('Mesh>Fire', fire_param)

constraint_param = {'V': str(s.compute.volume()), 
                    'max_iter': str(10000), 
                    'tol':str(1e-5)}
e.add_constraint('Mesh>Volume', constraint_param)

print('initial total energy:{}'.format(s.compute.energy(e)))
\end{lstlisting}

\begin{figure}[t!]
\centering
\includegraphics[width=\columnwidth]{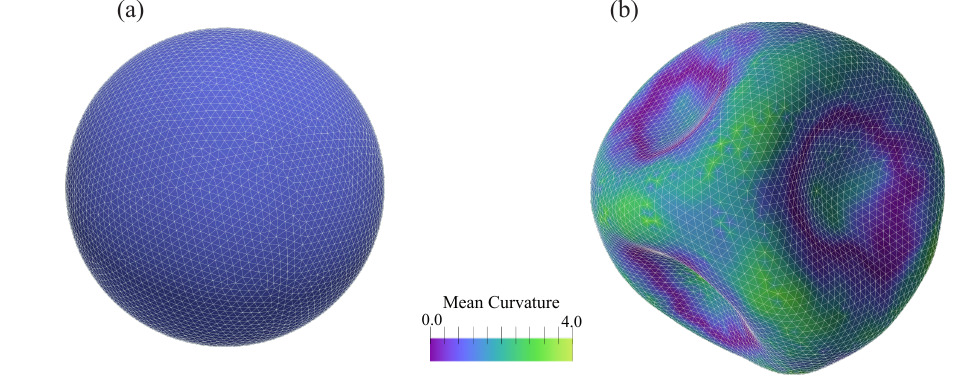}
\caption{(a) Initial configuration of a vesicle with unit radius comprising $N=6281$ vertices. The vesicle is subjected to a strain by increasing the edge rest length as $l_0 = (1+\epsilon) \left<l_e\right>$, where $\epsilon = 0.1$ and $\left<l_e\right>$ denotes the average edge length of the mesh. As a result, the vesicle buckles, leading to surface wrinkles, as depicted in (b). The colour bar indicates the local mean curvature. The final shape of the vesicle, after energy minimization under volume constraint, is obtained using the FIRE method~\cite{li2021chemically}.
}
  \label{fig:example_3}
\end{figure}

Finally, we can run the minimizer \lstinline|e.evolveMC(steps)| and visualize the results \lstinline|s.dumper.vtk()|. The initial and final configurations are shown in Fig. \ref{fig:example_3}. It's worth also noting that PyMembrane's flexibility is highlighted by the way parameters are passed using a dictionary which means that the user can conveniently load these values, e.g., from a JSON file.

\subsection{Periodic boundary condition}

\begin{figure}[!h]
  \includegraphics[width=\columnwidth]{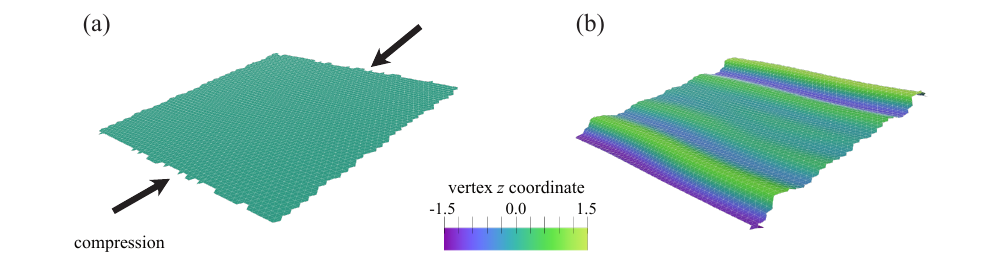}
  \caption{Brownian dynamics simulation for periodic thin elastic sheets that undergo wrinkling under uniaxial compression. The colour gradient represents the $z$ component of the vertex position, i.e., height from the reference $xy$ plane.}
  \label{fig:example_4}
\end{figure}

In many problems, probing finite-size effects is important, especially when modelling long-wavelength properties or when long-range interactions are present. A common approach is to use periodic boundary conditions. This method mimics an infinite system by replicating the simulation box, ensuring that particles that exit one side of the box re-enter from the opposite side.

PyMembrane offers a simple way to implement periodic boundary conditions. The simulation box is made periodic by setting the periodic flag to True via \lstinline|Box(..., periodic=True)|.

This example shows how to set up and run a simulation using periodic boundary conditions. We simulate wrinkling in a periodic thin sheet subject to uniaxial compression. To do this, we create an auxiliary function whose primary role is to refresh or update the box size:

\codecaption{Example of the function that updates the box size.}\label{code:code5}
\begin{lstlisting}[language=Python]
def compress_box(epsilon):
    new_box = Box(s.box.L.x*(1-epsilon), s.box.L.y, s.box.L.z, True)
    return new_box
\end{lstlisting}
Next, we study the dynamics by running a Brownian dynamics simulation as follows:
\codecaption{Executing Brownian dynamics simulation while compressing the box.}\label{code:code6}
\begin{lstlisting}[language=Python]
for _ in range(1, compress_steps):
    #evolve de system for run_steps
    e.evolveMD(steps=run_steps)
    #compress the box
    s.box = compress_box(strain_value)
\end{lstlisting}

The results are presented in Fig.~\ref{fig:example_4}. The periodic structure shows clear wrinkles. Notably, wrinkles emerge on the surface, indicative of the system's response to compression. 

\subsection{Bacterial Microcompartment}

\begin{figure}[!h]
  \includegraphics[width=\columnwidth]{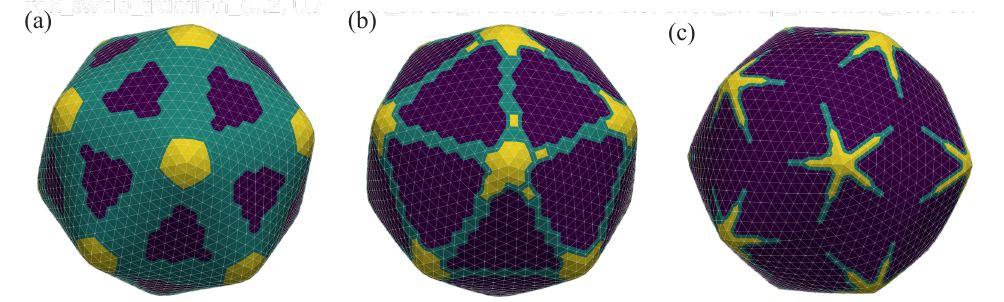}
  \caption{Morphology of multicomponent assemblies composed of three-vertex components. These components possess bending rigidities of 4.0 (blue), 1.0 (green), and 0.8 (yellow). They are presented in stoichiometric ratios of (a) $2$:$7$:$1$, (b) $6$:$3$:$1$, and (c) $8$:$1$:$1$. Studying such systems is crucial for understanding phenomena like bacterial microcompartments (BMCs). BMCs are self-assembling protein shells that facilitate the degradation of complex molecules, enabling bacteria to thrive in adverse conditions \cite{li2021effect}.}
  \label{fig:example_5}
\end{figure}

Bacterial microcompartments (BMCs) are protein shells encapsulating enzymes to enhance the metabolic process~\cite{yeates2008protein,kerfeld2018bacterial,kerfeld2005protein,yeates2010bacterial}. During assembly, pentameric, trimeric, and hexameric proteins aggregate, forming a buckled crystalline shell with a polyhedral shape~\cite{sutter2017assembly,lassila2014assembly,tanaka2010structure,Iancu2010}. To investigate the assembly, structure, and patterns, multi-scale models ranging from all-atom molecular dynamics (MD) simulations~\cite{li2021computational}, to coarse-grained MD simulations~\cite{mohajerani2018role,rotskoff2018robust}, and continuum models~\cite{vernizzi2011platonic,sknepnek2012buckling,li2021effect} have been utilized. Here, we reproduce the simulations of a continuum model and show how to generate and visualize BMC patterns and morphologies using PyMembrane.

In this example, we read from a mesh file named ``T192'', which is a closed shell mesh with icosahedral symmetry, where ``T'' is the triangulation number~\cite{caspar1962physical}. The radius of the shell is $R=11.5$ (in units of the average edge length), and the total number of vertices, edges and triangles are $10T+2$, $30T$, and $20T$, respectively. 

Next, we need to assign heterogeneous components on the shell to resemble the pentameric, trimeric and hexameric components of BMC. Here, we choose to tag vertices as being of three different types. Note that this procedure can also be done by assigning types to edges or triangles. A complete example using edge-based objects is included in PyMembrane examples, which reproduce the phase diagram in Ref.~\cite{li2021effect}. 

\codecaption{Assignment of heterogeneous BMC components to shell vertices.}\label{code:bmc2}
\begin{lstlisting}[language=Python]
vertices = system.getVertices()
Nv = len(vertices)
all_indices = set(range(Nv))

indices = [set(np.random.choice(list(all_indices), 
                                round(fac * Nv), 
                                replace=False))
           for fac in [fac0, fac1]]

indices.append(all_indices - indices[0] - indices[1])

for i, idx_set in enumerate(indices):
    for vi in idx_set:
        vertices[vi].type = i

system.setVertices(vertices)
\end{lstlisting}

After setting up the system, we create an instance of the Evolver class for Forces and Integrators as before. In this example, we use the same model for stretching energy as in Example~\ref{subsec:example1}, but another model for bending energy, the so-called Itzykson's discretization~\cite{kohyama2003budding}, 
\begin{equation}
E_\text{bend} = \sum_{i=1}^N A_i \left[\frac{1}{2}\kappa_{i,\alpha} \left(H_i-H_{i,0}\right)^2+\kappa^g_{i,\alpha} K_i\right],
\end{equation}
where $H_i$ and $K_i$ are the mean curvature and Gaussian curvature at vertex $i$, respectively. $H_{i,0}$, $\kappa_{i,\alpha}$, and $\kappa^g_{i,\alpha}$, respectively, correspond to the spontaneous curvature, the bending rigidity, and the saddle-splay modulus. Note that all these quantities can have values that depend on the vertex type.

In addition to the elastic and bending energy, the Line Tension potential is used to model the disaffinity between different types of vertices and the Limit potential is used to constraint the movement of each vertex so that the edge length lies between $l_{min}$ and $l_{max}$. After the setup of all the forces and parameters, we need to provide integrators to evolve the system. In this simulation, we use the Monte Carlo method to perform two types of movements: (1) $N_v$ vertex moves, with each vertex displaced by the distance $0.05l_0$, and (2) $N_v$ vertex id swaps, where the system randomly chooses two vertices and swap their type ids. This can be done by using the add$\textunderscore$integrator function. Here, we add \lstinline|Mesh>MonteCarlo>vertex>move| and \lstinline|Mesh>MonteCarlo>vertex>swap| integrators subsequently, which will be performed in turn.

\codecaption{Initialization of an instance of the Evolver class with force parameters and integrators. Multiple force types, such as Harmonic, BendingGK, Line Tension, and Limit, are added. Two integrators for Monte Carlo simulations are also defined.}\label{code:bmc3}
\begin{lstlisting}[language=Python]
harmonic_param = {'k': {'0': '10.0', '1': '10.0', '2': '10.0'},
                  'l0': {'0': '1.0', '1': '1.0', '2': '1.0'}}
e.add_force('Mesh>Harmonic', harmonic_param)

helfrich_param = {'kappaH': {'0': '4.0', '1': '1.0', '2': '0.8'},
                  'kappaG': {'0': '2.66668', '1': '-0.66667', 
                  '2': '-0.533336'}}
e.add_force('Mesh>BendingGK', helfrich_param)

ltension_param = {'gamma': {'0': '1.0', '1': '1.0', '2': '1.0'}}
e.add_force('Mesh>LineTension', ltension_param)

limit_param = {'lmax': {'0': '1.3', '1': '1.3'},
               'lmin': {'0': '0.7', '1': '0.7'}}
evolver.add_force('Mesh>Limit', limit_param)

e.add_integrator('Mesh>MonteCarlo>vertex>move', {'dr': '0.05'})
e.add_integrator('Mesh>MonteCarlo>vertex>swap', {'every step': '1'})
\end{lstlisting}

Next, we create a temperature list with ten temperatures in the range of $[10^{-1}, 10^{-7}]$ (measured in units of bending stiffness $\kappa$), where the highest temperature is chosen such that the MC acceptance ratio is around $80\%$. We run $2\times10^4$ steps at each temperature and dump the system state every $10^4$ steps. The snapshots of the simulation are captured using an instance of the Dump class, which we can produce output in the `vtk' format suitable for visualisation. Moreover, five cooling and reheating cycles are applied so that the system is fully relaxed. The visualized patterns are shown in Fig.~\ref{fig:example_5}, where the blue, green and yellow correspond to three components of BMC with soft, medium and rigid bending rigidities.

\codecaption{Running a simulated annealing simulation. Each of the ten temperatures involves $2\times10^4$ steps, with system states dumped every $10^4$ steps using an instance of the Dump class. The process includes five cooling and reheating cycles for system relaxation.}\label{code:bmc3}
\begin{lstlisting}[language=Python]
frame = 0
run_steps = 20000
snapshots = 10
anneal_steps = 5
templist = linspace(10**-1, 10**-7, 10)

for anneal in range(anneal_steps):
    for tempindex, temp in enumerate(templist):
        e.set_global_temperature(str(temp))
        for frame in range(snapshots):
            e.evolveMC(run_steps // snapshots)
            s.dumper.vtk(f"vertex_swap_t{frame}")
\end{lstlisting}

\section{Summary and Conclusions}

PyMembrane provides a wide range of models and tools for the rapid implementation and analysis of research-level simulations of liquid and elastic membranes. The high-level Python interface allows researchers to focus on the physics of the problem without the need to spend time on the technical details of the implementation. For those aiming to extend the package, a defining characteristic of PyMembrane is its modular design, guided by the principles of extensibility and adaptability. The software architecture, based on modern C++ coding practices, enables users to define forces, integrators, minimizers, and constraints with ease. This design philosophy facilitates the addition of new functionalities without compromising the overall layout of the package. 

\section{Future Directions}

Several enhancements to PyMembrane are slated for future releases: 
\begin{enumerate}[(1)]
    \item \textit{Parallel Processing:} Plans are underway to port the code to support multi-core processing using both MPI (Message Passing Interface) and OpenMP (Open Multi-Processing). This transition will greatly enhance the efficiency, enabling the software to handle larger and more complex simulations;
    \item \textit{GPU Integration:} Work is in progress to port PyMembrane to GPUs (Graphics Processing Units). By leveraging the parallel processing capabilities of modern GPUs, PyMembrane will benefit from significant speed-ups, especially for computationally intensive tasks (see, e.g.,~\cite{Matoz2020, Bahri2023ODDElastic});
    \item \textit{Extended Functionalities:} To provide a broader range of capabilities, there are plans to introduce additional force potentials and integrators. This expansion will make PyMembrane versatile for a more extensive set of simulations and modelling scenarios.
\end{enumerate}
The active community and open-source nature of PyMembrane promise a continuous evolution of the software, ensuring that it remains at the forefront of computational membrane modelling.

\section{Availability}

The PyMembrane code is released under the MIT licence and available on GitHub~\cite{matoz2023pymembrane}. The package is under active development and the GitHub repository provides both the latest version of the software and the collaborative environment for further improvements.

\section{Getting started with PyMembrane} 
PyMembrane is readily available on GitHub~\cite{matoz2023pymembrane}. We recommend starting with the detailed installation guide to ensure a smooth setup process. After installation, users can get acquainted with key features and functionalities by following some of the examples provided in this manuscript. These examples offer a hands-on introduction, aiding users in understanding the breadth and depth of what PyMembrane offers.

\section{Acknowledgments}

We thank Andjela Šarić for helpful discussions. R.S. acknowledges support from the UK EPSRC (Award EP/W023946/1). DMF was supported by the Comunidad de Madrid and the Complutense University of Madrid (Spain) through the Atracción de Talento program 2022-T1/TIC-24007. SL and MOdlC thank the support of the US Department of Energy (DOE), Office of Basic Energy Sciences under Contract DEFG02-08ER46539.  

\bibliographystyle{apsrev4-2} % uncomment this if you don't want titles
\bibliography{references}

\end{document}